\documentclass[11pt,a4paper]{article}
\pdfoutput=1
\usepackage{jheppub}
\usepackage[hyperref,svgnames]{xcolor}
\usepackage[latin1]{inputenc}
\usepackage{amsmath}
\usepackage{amsfonts}
\usepackage{amssymb}
\usepackage{booktabs}
\usepackage{graphicx}
\usepackage{setspace}
\usepackage{mathrsfs}
\usepackage{tikz}
\usepackage{caption}
\usepackage{subcaption}

\definecolor{hgreen}{rgb}{0,.3,0}
\definecolor{hred}{rgb}{.3,0,0}
\definecolor{hblue}{rgb}{0,0,.3}
\definecolor{LightGray}{gray}{0.95}

\numberwithin{equation}{section}

\title{Thin-Walled Higgs Assisted Q-balls from Pseudo-Nambu-Goldstone Bosons\\
}
\author[1]{Fady Bishara}
\emailAdd{fady.bishara@desy.de}

\author[2]{Olivier Lennon}
\emailAdd{olivier.lennon@physics.ox.ac.uk}

\affiliation[1]{Wolfgang-Pauli-Centre, Deutsches Elektronen-Synchrotron DESY, Notkestrasse 85, D-22607 Hamburg, Germany}
\affiliation[2]{Rudolf Peierls Centre for Theoretical Physics, University of Oxford, Clarendon Laboratory, Parks Road, Oxford OX1 3PU, United Kingdom}

\newcommand{\nf}{N_f}
\newcommand{\sun}[1]{{SU}(#1)}

\newcommand{\un}[1]{{U}(#1)}

\newcommand{\avg}[1]{\mathrm{tr} \left( #1\right)}
\newcommand{\delsig}[2]{\partial_{#1}^{#2}\Sigma}
\newcommand{\delsighc}[2]{\partial_{#1}^{#2}\Sigma^\dagger}
\newcommand{\sighc}{\Sigma^\dagger}

\newcommand{\mcL}{\mathcal{L}}

\newcommand{\mbbH}{H}
\newcommand{\mbbS}{S}

\newcommand{\VEV}{v_h}

\newcommand{\mpi}{m_\pi}

\newcommand{\lamp}{\lambda_p}
\newcommand{\lams}{\lambda_s}
\newcommand{\lamh}{\lambda_h}

\newcommand{\lamchihs}{\Lambda_\chi^\textsc{hs}}

\newcommand{\nh}{\eta}

\abstract{
We consider the question of whether Q-balls can exist in a chiral Lagrangian truncated at leading order when, in addition, the Standard Model Higgs boson couples to the pseudo-Nambu-Golstone bosons (pNGBs). In particular, we consider the so-called thin-wall limit where volume energy dominates over surface energy. It is known that the leading order chiral Lagrangian alone does not support such multi-field solutions. Augmented by the Higgs, however, we do indeed find that such solutions exist. We then study their properties numerically and, in various limits, analytically.
Furthermore, since we consider a mirror-world-like model where the pNGBs are composite states of fundamental fermions, the question of Fermi repulsion in the high density bulk of the Q-ball plays a central role in determining its properties.
The main effect is that when the parameter controlling the Fermi repulsion increases beyond a critical value, the radius of the Q-ball increase and continues to increase while the Q-ball becomes more weakly bound. 
As a result, there are Q-ball solutions with radii well exceeding a femtometer which would interact with nuclei in direct detection experiments via momentum-dependent form factors making their signatures striking. We leave the question of the production and direct detection of these Q-balls to a future study.
}

\graphicspath{{./figs/}}

\preprint{OUTP-21-22P, DESY-21-155}

\begin{document}

\tikzstyle{every picture}+=[remember picture]
\usetikzlibrary{shapes.geometric}
\usetikzlibrary{calc}
\usetikzlibrary{decorations.pathreplacing}
\usetikzlibrary{decorations.markings}
\usetikzlibrary{decorations.text}
\usetikzlibrary{patterns}
\usetikzlibrary{backgrounds}
\usetikzlibrary{positioning}
\tikzstyle arrowstyle=[scale=2]
\tikzstyle directed=[postaction={decorate,decoration={markings,
		mark=at position 0.6 with {\arrow[arrowstyle]{>}}}}]
\tikzstyle rarrow=[postaction={decorate,decoration={markings,
		mark=at position 0.999 with {\arrow[arrowstyle]{>}}}}]

\everymath{\displaystyle}

\maketitle

\section{Introduction}

Q-balls are non-topological solitons~\cite{Lee:1991ax} whose stability is ensured, simultaneously, by energy and Noether charge conservation~\cite{Coleman:1985ki}.
Specifically, a Q-ball is the most energetically-favourable state that stores charge in a theory.
In his classic analysis~\cite{Coleman:1985ki}, Coleman considered a single complex scalar field that carries a global $U(1)$ charge.
The resulting extended objects have piqued theorists' interests since then, as evidenced by the large body of work that ensued.\footnote{Coleman's paper~\cite{Coleman:1985ki} alone has, at the time of writing, greater than $800$ citations.} 
The original Q-ball analysis has thus been expanded to accommodate multiple fields~\cite{Kusenko:1997zq}, more complicated symmetry groups~\cite{Safian:1987pr}, as well as gauging the stabilising symmetry~\cite{Lee:1988ag, Heeck:2021zvk}.

Finding the extremum, in field space, that corresponds to a minimum of the energy functional of the soliton can be formulated by studying a bounce equation, albeit at fixed and finite charge. This latter constraint is most naturally implemented with a Lagrange multiplier~\cite{Kusenko:1997si}.
Consequently, an exact analytical solution is generally not possible and one must resort to numerical techniques.
However, exact analytical solutions can be found in certain limits such as the `thin-wall' limit~\cite{Coleman:1985ki}, the `thick-wall' limit~\cite{Kusenko:1997ad}, and, more recently even beyond the strict thin-wall limit~\cite{Heeck:2020bau}.
More precisely, thin-walled Q-balls are spherically-symmetric solutions with parametrically large volumes such that the energy of the Q-ball is dominated by the bulk, or volume, energy rather than the surface energy.\footnote{By parametrically large, we mean here a volume per unit charge greater than the corresponding Compton wavelength of the constituent scalar.}
Conversely, thick-walled Q-balls are appropriate to describe small Q-balls when the surface energy gives an important contribution to the overall rest mass of the object.

An interesting question to ask is whether the Standard Model (SM) itself, with its known field content, can admit Q-ball solutions, especially since the SM has many accidental global symmetries.
While only the matter fields are charged under these global symmetries, however, they are inherited by the scalar composite states they form below the QCD confinement scale.
This question was addressed in~\cite{Distler:1986ta} where the authors considered the leading order $SU(3)$-flavour-symmetric chiral Lagrangian.
The Q-balls in that study were envisioned to be composed of kaons which are stable in the limit of strangeness conservation.
It turns out, however, that no Q-ball solutions exist at leading order.

Nevertheless, Q-balls arise in many extensions of the SM, such as supersymmetric or mirror-world-like theories where additional scalar fields and global symmetries are aplenty~\cite{Kusenko:1997si, Kusenko:1997zq, Bishara:2017otb}, or in theories of extra dimensions~\cite{Demir:2000gj, Abel:2015tca}. They have long been considered as candidates for dark matter~\cite{Kusenko:1997si, Kusenko:2001vu, Graham:2015apa, Ponton:2019hux} owing to their stability against decay.
Q-balls can even play a role in baryogenesis~\cite{Krylov:2013qe}.
So far, the existence of Q-balls has not been confirmed experimentally, but, should they cross-paths with a sensitive detector, their signatures will be striking~\cite{Gelmini:2002ez, Kusenko:1997vp, Croon:2019rqu}.

In this work, we focus on a class of multi-field, thin-walled Q-balls that are stabilised by a global $U(1)$ symmetry in a theory beyond the SM.
We have in mind a mirror-world-like scenario~\cite{Kobzarev:1966qya, Foot:1991bp, Foot:2014mia}, but leave the analogue of $U(1)_Y$ ungauged. The leading order interaction between this hidden sector and the Standard Model is given through a portal interaction between the SM Higgs and a new scalar multiplet that acquires a vacuum expectation value (VEV).
Upon spontaneous symmetry breaking, the matter fields of the hidden sector acquire masses via Yukawa terms with the scalar.
The resulting breaking of an approximate chiral symmetry in the sector of hidden ``quarks'' leads to a number of stable scalars. The couplings of these pseudo-Nambu-Goldstone bosons (pNGBs) with the SM Higgs provide a non-trivial potential, as seen in Eq.~\eqref{eq:lag-higgs-coupling-with-higgs}, for which it is natural to ask if stable Q-ball states exist.
Here, the pNGBs are composite states of fundamental fermions and, therefore, at high density, Fermi repulsion must be taken into account as it modifies both the mass and volume of the Q-ball.

This scenario was explored in the thick-wall Q-ball limit in previous work~\cite{Bishara:2017otb}. The nature of the thick-wall limit is that the fields take on moderately low values in the Q-ball, and so higher order terms can be neglected in the Hamiltonian.
However, this is not the case in the thin-wall limit.
Higher order terms could render the potential unfavourable to Q-ball solutions.

The goal of this work is to show that Q-balls in the thin-wall limit of this theory exist and are classically stable against decay to the pions of the hidden sector.
We present numerical results, considering the effect of Fermi repulsion, as well as analytic expressions in certain limits.
These large Q-balls are interesting phenomenologically as they could form a component of dark matter.
Owing to their interactions with Standard Model states via the Higgs portal they can be detected in direct detection experiments.
Their signatures, however, are different to those of ordinary point particles due to form factor suppression at moderately-high momentum transfer~\cite{Gelmini:2002ez}. We leave this to future work on the phenomenology of this class of theories.

Note that while we have a particular UV model in mind, the only relevant states are the low energy ones, namely, the pNGBs and the SM higgs. The dynamics of the pNGBs are fully determined by the low energy effective field theory (EFT) according to the Callan-Coleman-Wess-Zumino
coset construction~\cite{Coleman:1969sm,Callan:1969sn} and, therefore, our analysis is more general than the model we describe. The only caveat is that in the model we have in mind, the couplings of the Higgs to the hidden sector pions are fixed by the breaking of scale invariance~\cite{Voloshin:1980zf,Voloshin:1985tc,Chivukula:1989ds} in addition to the portal coupling.

This paper is structured as follows. In Section~\ref{structure}, we describe the structure of the hidden sector of study. We introduce the Higgs portal term that leads to the coupling of the Standard Model Higgs with the pseudo-Nambu-Goldstone bosons that arise from the spontaneous symmetry breaking of approximate chiral symmetry in the ``quark'' sector of the theory. In Section~\ref{main}, we construct the Q-ball solution by analysing the scalar sector for states that minimise the energy for a given charge. We then present our numerical results, as well as regions of parameter space that allow for an analytic description. We finally discuss constraints on these solutions.

\section{The structure of the model}
\label{structure}

In this section, we briefly outline the model we later analyse for thin-wall Q-balls. The model has similarities with Mirror World~\cite{Kobzarev:1966qya, Foot:1991bp, Foot:2014mia} and Twin Higgs~\cite{Chacko:2005pe,Chacko:2005vw,Chacko:2005un} scenarios, and in particular the Fraternal Twin Higgs models~\cite{Craig:2015pha,Garcia:2015loa,Garcia:2015toa,Craig:2015xla,Farina:2016ndq}. In our case, however, we are taking the hidden sector $\sun{3}'$ dynamical scale $\lamchihs\gtrsim1\,\mathrm{TeV}$ rather than the few GeV appropriate for the Twin Higgs models.

\subsection{The Content of the Hidden Sector}

In addition to the SM field content, we consider an $SU(3)'\times SU(2)'$ gauge group in the hidden sector (HS) leaving the analogue of $U(1)_Y$ ungauged.
The $SU(3)'$ sector is QCD-like, i.e., it is asymptotically free in the UV and
confines at energies below a scale $\Lambda_\chi^{\mathrm{HS}}$.

As for the $SU(2)'$ gauge group, we assume that it is spontaneously broken by a VEV of a scalar doublet, $S$. 
A Higgs portal coupling between $S$ and the SM doublet, $H$, gives the leading interaction between the two sectors.
We consider the following potential, which facilitates both spontaneous symmetry breaking in both sectors, as well as mass eigenstate mixing:
\begin{equation}
V(\mbbH,\mbbS) = -\mu_h^2 \mbbH^\dagger \mbbH + \lamh(\mbbH^\dagger \mbbH)^2 - \mu_s^2 \mbbS^\dagger\mbbS + \lams(\mbbS^\dagger\mbbS)^2
+ \lamp(\mbbH^\dagger\mbbH) (\mbbS^\dagger\mbbS)\,.
\label{eq:scalar-pot}
\end{equation}
The final term above results in the mixing of the SM and HS Higgs gauge eigenstates $h'$ and $s'$ into the mass eigenstates $h$ and $s$. This dimension-four operator is consistent with the symmetries of the theory, and could arise from integrating out higher modes.

We can determine the masses of the corresponding eigenstates of this theory. Writing the VEVs in the two sectors as $\left<H\right> = v_h / \sqrt{2}$ and $\left<S\right> = v_s / \sqrt{2}$,
the gauge eigenstate $s^\prime$ can be written in terms of the mass eigenstates as $s' \approx s - \theta h$,
where the mixing angle $\theta$, in the small angle approximation, is given by,
\begin{equation}
\label{eq:theta-definition}
\theta \approx \frac{\lambda_p v_h}{2 \lambda_s v_s}.
\end{equation}
The quartic Higgs coupling in the same approximation is, then, $\lambda \approx \lambda_h - \lambda_p^2/(4\lambda_s)$,
where $\lambda_h$ is its SM value; for more details regarding the scalar mass matrix diagonalization see~\cite{Barger:2008jx} and Appendix~A in~\cite{Bishara:2017otb}.

Minimally, we mirror one full SM matter generation in the same representations and with the same charges under $SU(3)' \times SU(2)'$.
These states acquire masses through Yukawa couplings to the HS scalar doublet, S:
\begin{equation}
\mathcal{L}^{\text{HS}}\supset y_{ij}\overline{Q}_{L,i}Sq_{R,j}+\text{h.c.},
\label{eq:yukawa}
\end{equation}
For concreteness, we assume that the resulting quark masses are `light', i.e., below $\Lambda_\chi^{\mathrm{HS}}$, while the lepton masses are heavy. 
These `light' quarks will hadronise into stable pseudo-Nambu-Goldstone bosons (pNGBs).
Allowing for an arbitrary number of `heavy' quark generations contributes to the HS scalar to pion couplings as in the SM via matching onto an $S\langle G\cdot G\rangle$ operator~\cite{Chivukula:1989ds}.

\begin{figure}\centering
	\begin{tikzpicture}[line cap=round,
	decoration={brace,amplitude=7pt}]
		\def\dx		{1.0};	\def\ddx	{0.1};	\def\VEVh	{-2.25};
		\def\VEVs	{-0.4};	\def\lamx	{0.0};	\def\mpi	{-1.25};
		\def\dheavy	{0.6}
		\node  [anchor=west] at (\dx+\ddx,\VEVs) {$m_s$};
		\node  [anchor=west] at (\dx+\ddx,\lamx) {$\Lambda_\chi^\text{\sc hs}$};
		\node  [anchor=west] at (\dx+\ddx,\mpi) {HS pions $\sim\mathcal{O}$(few TeV)};
		\node  [anchor=west] at (\dx+\ddx,\VEVh) {$m_h\sim v_h$};
		\node  [anchor=south] at (0,\lamx+1) {$\vdots$};
		\draw [thick] (\dx,\VEVs) -- (-\dx,\VEVs);
		\draw [thick] (\dx,\lamx+0.6 ) -- (-\dx,\lamx+0.6 );
		\draw [thick] (\dx,\lamx+0.75 ) -- (-\dx,\lamx+0.75 );
		\draw [ultra thick, line cap=round] (\dx,\lamx) -- (-\dx,\lamx);
		\draw [thick] (\dx,\mpi) -- (-\dx,\mpi);
		\draw [thick] (\dx,\mpi+0.1) -- (-\dx,\mpi+0.1);
		\draw [thick] (\dx,\mpi-0.1) -- (-\dx,\mpi-0.1);
		\draw [thick] (\dx,\VEVh) -- (-\dx,\VEVh);
  \draw [decorate,very thick,align=left] (\dx+0.3,\lamx+\dheavy+1.1) -- (\dx+0.3,\lamx+\dheavy)
  node [midway,anchor=west,inner sep=2pt, outer sep=10pt]{heavy HS mesons\\and baryons};
	\end{tikzpicture}
	\caption{Spectrum of the hidden sector states relative to the weak scale $v_h$. In our analysis, we assume that the hidden-sector Higgs boson, $s$, is close to or above $\Lambda_\chi^\textsc{hs}$ and thus we do not consider it as a low-energy degree-of-freedom in our EFT.}
	\label{fig:SpectrumHS}
\end{figure}
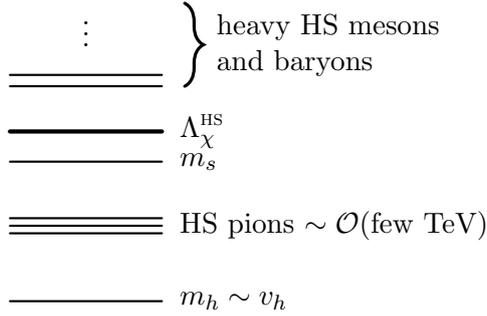

The spectrum of this theory is shown in Fig.~\ref{fig:SpectrumHS}. Note that we assume that the mass of the HS Higgs, $m_s$, to be larger than the HS pion masses in our theory such that it is not a relevant degree of freedom in the low energy EFT. It can, however, be above or below the chiral symmetry breaking scale, $\Lambda_\chi^{\textsc{hs}}$.

\subsection{Chiral Symmetry Breaking in the Hidden Sector}

A necessary condition for the existence of Q-ball solutions is that the relative binding energy of the Q-ball per unit charge, $1-M_Q/Q m_\pi$, be greater than zero. Furthermore, in the case of thin-wall Q-balls, the fractional binding energy scales as $\epsilon$  with $\epsilon\ll 1$ and thus the relevant degrees of freedom are indeed the pNGBs whose dynamics are described by a low-energy chiral Lagrangian.
The global flavour symmetry is spontaneously broken to its diagonal subgroup, $SU(n_l)_L \times SU(n_l)_R\to SU(n_l)_V$, and explicitly broken by the Yukawa terms.\footnote{As in Ref.~\cite{Bishara:2017otb}, we ignore the fact that the symmetry group is generally $\un{\nf}_L\times \un{\nf}_R$ since the one non-anomalous $U(1)$ from the $\un{\nf}_L\times \un{\nf}_R$, that in the SM case corresponds to baryon number, acts trivially on the pNGBs, so it is not of interest to us here.} 
Provided that the mass matrix $M$ is not proportional to the unit matrix, the remaining global symmetry acting on the pNGBs is $U(1)^{n_l - 1}$, in the absence of other interactions. Under the assumption that the number of light quarks is 2, by Goldstone's theorem there will be $3$ pNGBs: the HS pions, which transform under the unbroken $U(1)$ symmetry. Below the chiral symmetry breaking scale of the HS, the correct effective description is the chiral Lagrangian in terms of these pNGBs.

To describe the pion sector of the theory, we conventionally define the unitary matrix field of unit determinant from the three pNGBs, $\pi^a$,
\begin{equation}
\Sigma = \exp (i \pi^a T^a / f),
\end{equation}
where $T^a$ are the generators of $SU(n_l)$ and $f$ is a coefficient to be determined experimentally. Under the global vectorial symmetry, $\Sigma$ transforms as
\begin{equation}
\Sigma \to \Sigma' = V \Sigma V^{\dagger},
\end{equation}
where $V$ is given in terms of the Hermitian and traceless matrix $X$ as $V = \exp(-iX)$. The leading order chiral Lagrangian is then
\begin{equation}
\mcL = \dfrac{f^2}{4} \avg{\delsig{\mu}{}\delsighc{}{\mu}} + \dfrac{B_0 f^2}{2}\avg{M(\Sigma + \sighc - 2)}.
\end{equation}
where $B_0$ is another coefficient to be determined experimentally and $M$ is the quark mass matrix.
This is the leading-order low-energy description of the HS pions, valid up to the chiral symmetry breaking scale $\Lambda^{\mathrm{HS}}_\chi \sim 4\pi f$. Despite the non-trivial potential contained within this Lagrangian, it does not admit thin-wall Q-ball solutions~\cite{Distler:1986ta}.

The HS pions of the theory are not the only light scalars which must be taken into consideration. The lightest mass eigenstate, $h$, associated to the scalar doublets couples to the chiral Lagrangian in a manner completely determined through the breaking of scale symmetry~\cite{Voloshin:1980zf,Voloshin:1985tc,Chivukula:1989ds}.
It is relevant to our analysis provided that its mass is smaller than those of the HS pions. Specifically, the leading order coupling of the gauge eigenstate, $s'$, to the HS chiral Lagrangian is given through
\begin{equation}
\mcL = \left(1+ \frac{4 n_h}{3 \beta_0} \frac{s'}{v_s} \right)\dfrac{f^2}{4} \avg{\delsig{\mu}{}\delsighc{}{\mu}} + \left(1 + \left[1+ \frac{2 n_h}{\beta_0}\right] \frac{s'}{v_s} \right) \dfrac{B_0 f^2}{2}\avg{M(\Sigma + \sighc - 2)},
\label{eq:chiral-lag}
\end{equation}
where $n_h$ is the number of quarks with mass greater than the confinement scale, and $\beta_0$ is the one-loop beta function. In principle, higher powers of $s'/v_s$ can also be included in Eq.~\eqref{eq:chiral-lag}. However, the resulting couplings to the light mass eigenstate, $h$, are suppressed by commensurate powers of the small mixing parameter $\theta$. Therefore we are justified in ignoring such terms. These additional terms originate from the Yukawa couplings of the fundamental quarks with the scalar $S$, and from integrating out the heavy quark degrees of freedom. The details of the numerical coefficients can be found in Ref.~\cite{Chivukula:1989ds}.

We assume that the heavier scalar, $s$, has a mass larger than the chiral symmetry breaking scale, and is thus irrelevant for this low-energy description. The full low-energy Lagrangian describing the HS scalars with masses less than $\Lambda^{\mathrm{HS}}_\chi$ is then
\begin{equation}
\begin{split}
\mcL = 
&\left(1-\theta\dfrac{2\nh}{3}\dfrac{h}{v_s}\right)\dfrac{f^2}{4}\avg{\delsig{\mu}{}\delsighc{}{\mu}}
+ \left(1-\theta\left(1+\nh\right)\dfrac{h}{v_s}\right)
\dfrac{B_0 f^2}{2}\avg{M(\Sigma + \sighc - 2)}\\
&+ \dfrac{1}{2}\partial_\mu h \partial^\mu h - U(h),
\label{eq:lag-higgs-coupling-with-higgs}
\end{split}
\end{equation}
where we have defined $\eta = 2 n_h / \beta_0$ and the Higgs potential is given by
\begin{equation}
U(h)=\dfrac{1}{2}m_h^2 h^2 +\lambda \VEV h^3 + \frac{1}{4}\lambda h^4.
\end{equation}

\section{Higgs assisted thin-wall Q-balls}
\label{main}

The hidden sector theory with a Higgs portal to the SM described in the previous section was shown to admit thick-wall Q-ball solutions in Ref.~\cite{Bishara:2017otb}. In this paper, we turn our attention to the question of whether it admits \emph{thin-wall} Q-ball solutions, where the volume energy dominates and the surface energy is negligible.

\subsection{Minimising the Energy in a Sector of Fixed Charge}

A Q-ball is a solution of minimum energy at fixed charge. The energy functional that should be minimized is given by
\begin{equation}
\mathcal{E}_{\omega}=H+\mathcal{E}_F+\omega\left(Q - \int\mathrm{d}^3x \, J^0 \right),
\label{eq:eulerian}
\end{equation}
where $H$ is the Hamiltonian of the theory with Lagrange density given by Eq.~\eqref{eq:lag-higgs-coupling-with-higgs} and $\omega$ is a Lagrange multiplier that enforces the fixed charge constraint.
The second term on the right hand side, $\mathcal{E}_F$, is the total Fermi repulsion energy which must be included to account for the fact that the HS pions are fermion/anti-fermion composites states.
Therefore, the overlap between HS pion wave functions cannot be arbitrarily large.

The average energy contributed to the Q-ball per constituent fermion is
\begin{equation}
\frac{3}{5}E_F = \frac{3}{10m_f} (3 \pi^2 n)^{2/3}\,,
\end{equation}
where $E_F$ is the Fermi energy, $n$ is the number density of a fermionic species, and $m_f$ is the dressed quark mass, i.e., the mass of an excitation  with the same quantum numbers as a quark, within the Q-ball medium. Typically, $m_f \sim \Lambda^{\mathrm{HS}}_{\chi}$, which is not set in our theory and so can in principle be large relative to other scales.\footnote{If the chiral symmetry breaking scale is sufficiently high, one might wonder if it might give rise to unnaturally large corrections to the Higgs mass through pion loops. The cubic Higgs-pion coupling gives rise to corrections that are logarithmic in $\lamchihs/\mpi$, however, and therefore naturalness is not a problem in this case.} The total energy contributed to the Q-ball is thus
\begin{equation}
\mathcal{E}_F = \frac{1}{5 m_f}\left(243 \pi^4 \frac{Q^5}{V^2}\right)^{1/3},
\end{equation}
where $V$ is the volume of the Q-ball, and $Q$ the total charge.

The Noether charge (last term in Eq.~\eqref{eq:eulerian}) is associated to the invariance of the Lagrangian in Eq.~\eqref{eq:lag-higgs-coupling-with-higgs} under the transformation
\begin{equation}
\Sigma \to \exp(-i\alpha X)\Sigma\exp(i\alpha X) \quad \mathrm{and} \quad h\to h,
\end{equation}
where $X$ is a Hermitian charge operator. 
The Noether charge functional is given by
\begin{equation}
\label{eq:ChargeFunc}
\int\mathrm{d}^3x \, J^0 = i\int\mathrm{d}^3 x \left(1-\theta\dfrac{2\nh}{3}\dfrac{h}{v_s}\right)\dfrac{f^2}{4}\avg{\dot{\Sigma}[\Sigma,X] + \dot{\Sigma}^\dagger [\Sigma^\dagger, X]}.
\end{equation}
Hence, a Q-ball must be time-dependent solution of the classical equations of motion as is well known~\cite{Coleman:1985ki}. Explicitly isolating the time-dependent terms in the energy functional gives
\begin{equation}
\begin{alignedat}{2}
\mathcal{E}_\omega = & \int\mathrm{d}^3 x && 
\left\{\left(1-\theta\dfrac{2\nh}{3}\dfrac{h}{v_s}\right)\dfrac{f^2}{4}\avg{|\dot{\Sigma}-i\omega[\Sigma,X]|^2 - \omega^2[\Sigma,X][X,\Sigma^\dagger] + \nabla\Sigma\cdot\nabla\Sigma^\dagger}\right.\\
&	&& \kern-2.5em \left. - \left(1-\theta\left(1+\nh\right)\dfrac{h}{v_s}\right)
\dfrac{B_0 f^2}{2}\avg{M(\Sigma + \sighc - 2)}\right.\\
&	&& \kern-2.5em  \left. + \dfrac{1}{2}\left(\dot{h}^2+\nabla h\cdot\nabla h\right)+ U(h)\right\}+\omega Q + \mathcal{E}_F\,.
\end{alignedat}
\end{equation}
The two terms with explicit time-dependence are minimised if they vanish,\footnote{The field configuration $\Sigma(\vec{r},t) = 1$ leads to $Q=0$. Thus, a configuration with non-zero charge must differ from the vacuum in some domain.} and so
\begin{equation}
\label{eq:QballAnsatz}
\Sigma(\vec{r},t) = \exp(-i\omega Xt)\Sigma(\vec{r})\exp(i\omega Xt) \quad \mathrm{and} \quad h(\vec{r},t) = h(\vec{r})\,.
\end{equation}

After reinsertion of the Q-ball ansatz into the energy functional, what remains is known as a Euclidean bounce~\cite{Coleman:1977py, Callan:1977pt, Coleman:1977th}. This well-studied class of differential equations is generally intractable analytically, but progress can be made in certain limits. In previous work, the thick-wall limit~\cite{Kusenko:1997ad} was analysed for Q-ball solutions~\cite{Bishara:2017otb}. This work is concerned with the opposite limit: thin-wall Q-balls~\cite{Coleman:1985ki}.

A thin-wall Q-ball is characterised by a core of a homogeneous state, named Q-matter, and a thin outer shell. The mass of a thin-wall Q-ball is dominated by this core.\footnote{This statement holds apart from in the case where the underlying fields comprising the Q-ball take on configurations such that the potential energy vanishes inside the homogeneous core (see Ref.~\cite{Spector:1987ag}). In this case, the mass of the resulting Q-ball is dependent only on its surface energy.} We let $\Sigma(\vec{r})=\Sigma_0$ and $h(\vec{r})=h_0$ be constant spatial profiles of the fields inside the core of the Q-ball, such that
\begin{equation}
\label{eq:HomogEul}
\begin{split}
\mathcal{E}_\omega \approx & - \omega^2 \left(1-\theta\dfrac{2\nh}{3}\dfrac{h_0}{v_s}\right)\dfrac{f^2}{4}\avg{[\Sigma_0,X][X,\Sigma_0^\dagger]} V \\
& - \left(1-\theta\left(1+\nh\right)\dfrac{h_0}{v_s}\right)
\dfrac{B_0 f^2}{2}\avg{M(\Sigma_0 + \sighc_0 - 2)}V \\
& + U(h_0)V + \omega Q + \mathcal{E}_F,
\end{split}
\end{equation}
where $V$ is the volume of the core of the Q-ball.
To determine the mass of the resulting Q-ball, this expression must be minimised with respect to the field content, as well as the volume and the Lagrange multiplier.

So far, we have not specified the number of light flavours, $n_l$, in the analysis that led to Eq.~\eqref{eq:HomogEul}. To make further progress, it is useful to specialize to our case of interest, $SU(n_l)=SU(2)$, which will allow us to simplify the exponentials involving $\hat n \cdot \vec{T}$, where $T^a$ are the generators of $SU(n_l)$.
In this case, $T^a\propto \sigma^a$, where $\sigma^a$ are the Pauli matrices.
Since $\Sigma_0$ is now an element of $SU(2)$, we can write it as, 
\begin{equation}
\Sigma_0=\exp\left(i\varphi\hat{n}\cdot\sigma\right)=\cos\varphi+i\left(\hat{n}\cdot\sigma\right)\sin\varphi,
\end{equation}
where it is understood that $\cos\varphi$ multiplies a unit $2\times2$ matrix, and $\hat{n} = (n_1,\, n_2,\, n_3)$ is a unit vector, $\hat{n}^2 = 1$.
Furthermore, without loss of generality, we can choose the charge operator, which appears in Eq.~\eqref{eq:QballAnsatz}, $X=\sigma_3/2$ since the Lagrangian is invariant under $SU(2)$ transformations.

For convenience, we introduce a small parameter, $\epsilon$, defined as,
\begin{equation}
\epsilon \equiv \theta \dfrac{h_0}{v_s}\,.
\end{equation}
Here, we recall that $\theta$ is the mixing angle between the SM and HS Higgses and is less than unity following our choice to work in the small angle limit. Furthermore, the ratio of VEVs $h_0/v_s$ must also be small otherwise the effective field theory describing our system is not valid. Therefore, we also have that $\epsilon\ll 1$.
Minimising with respect to the Lagrange multiplier, $\omega$, yields
\begin{equation}
\label{eq:ChargeFunctional}
Q = \left(1-\dfrac{2\nh}{3}\epsilon\right) f^2 \omega(1-n_3^2)\sin^2\varphi V\,.
\end{equation}
This expression corresponds precisely to the one for the charge as determined from Eq.~\eqref{eq:ChargeFunc}. We use this to eliminate $\omega$, giving \begin{equation}
\begin{split}
\label{eq:GeneralQ-ballEnergy}
E = & \dfrac{Q^2}{2\left(1-\dfrac{2\nh}{3}\epsilon\right) f^2 (1-n_3^2)\sin^2\varphi V}
+ \left(1-\left(1+\nh\right)\epsilon\right) \mpi^2 f^2 (1 - \cos\varphi) V \\
& + U(h_0)\, V + \frac{1}{5 m_f}\left(243 \pi^4 \frac{Q^5}{V^2}\right)^{1/3}.
\end{split}
\end{equation}
The only dependence on the direction of the VEV of $\Sigma$ is the factor $n_3$ in the first term. The energy of the Q-ball is minimized for $n_3=0$, which corresponds to zero VEV for the neutral pions. This behaviour is unsurprising since a neutral pion VEV inside the Q-ball contributes to its mass but not to its charge -- unlike the SM Higgs, the neutral pion does not offer a way to reduce the mass of the resulting Q-ball for a given charge.

We further note that, in the expression that must be minimised, $E/Q$, the volume and charge of the Q-ball always appear in the form $V/Q$. Thus, we infer that $V$ scales linearly with $Q$ for thin-wall Q-balls, even in the presence of Fermi degeneracy pressure. This is in contrast with the thick-wall case, where $V \sim Q^{-3}$, for small $Q$~\cite{Kusenko:1997ad, Bishara:2017otb}.

Before we continue, we note the complications in proceeding analytically. Due to the non-trivial dependence on $V$, it is not feasible to analytically minimise the energy of the Q-ball with respect to the volume exactly. This is entirely due to the presence of the Fermi degeneracy pressure.
Furthermore, the presence of $n_h$ heavy quarks in Eq.~\eqref{eq:GeneralQ-ballEnergy} does not qualitatively affect the Q-ball solution since it only contributes an additional (and slightly larger) coupling between the HS Higgs and pions. Therefore, in the following, we will set $n_h=0$ to simplify our expressions -- we will reintroduce this parameter in our numerical results. The Higgs potential is also a barrier to obtaining exact analytical expressions. We will discuss this as we proceed below.

\subsection{Higgs-Assisted Thin-Wall Q-balls}

For notational convenience, let us define the dimensionless variables,
\begin{eqnarray}
\hat{E}\equiv \frac{E}{m_\pi\,Q}\,,\qquad\text{and}\qquad \nu\equiv \frac{m_\pi\,f^2\,V}{Q}\,,
\end{eqnarray}
together with the dimensionless parameters
\begin{equation}
\boxed{A \equiv \frac{(243\pi^4)^{1/3}}{5}\left(\frac{f^4}{m_f^3\,m_\pi}\right)^{1/3} \quad \mathrm{and} \quad B  \equiv \frac{1}{4}\frac{\lambda}{\theta^4}\frac{v_s^4}{m_\pi^2\,f^2} \quad \mathrm{and} \quad C \equiv \theta\frac{v_h}{v_s}.}
\label{eq:dimless-pars}
\end{equation}
This equation is enclosed in a box for referential convenience since all of the ensuing analysis will be done using these dimensionless parameters defined here. These parameters effectively control the size of the contribution to the Q-ball mass due to the presence of Fermi degeneracy pressure, and the SM Higgs potential. Note that $C$ is not entirely independent of $B$; they are related through
\begin{equation}
\label{eq:BC4}
BC^4 = \frac{\lambda}{4} \frac{v_h^4}{m_{\pi}^2 f^2}.
\end{equation}
This relation is always satisfied for some parameter values in each theory. Moreover, by Eq.~\eqref{eq:theta-definition}, we see that $C \propto \theta^2$, and so it is related to the mixing between the SM and the HS. Thus, we see that $C$ is naturally small.

Consider the example of $m_\pi \sim f \sim 4v_h$ and $\lambda \sim 0.1$, then $ B C^4 \sim 10^{-4}$. Since, in the small angle approximation, $\theta \lesssim 10^{-1}$, we also have that
\begin{equation}
B \gtrsim 10^3 \left( \frac{v_s}{f} \right)^4, \quad C \lesssim 10^{-1} \left(\frac{v_h}{v_s}\right), \quad \frac{\lambda_p}{\lambda_s} \left(\frac{v_h}{v_s}\right) \lesssim 10^{-1},
\end{equation}
where the final inequality comes from the definition of $\theta$ in Eq.~\eqref{eq:theta-definition}. If there is no hierarchy between the two parameters $\lambda_p$ and $\lambda_s$, then we see that $C \lesssim 10^{-2}$ and $B \gtrsim 10^4$, which in turn requires that $v_s \gtrsim f$ -- this latter constraint is consistent with the idea that the ``heavy Higgs'' is massive enough that we need not include it in our effective theory. Realistically, the parameters can be different from these assignments, but we use values similar to these in our numerical study below.

Q-ball stability requires, at the minimum of the energy, that $\hat{E}<1$; that is, the mass of the Q-ball must be less than the product of the total charge and the HS pion mass such that it cannot classically decay into pions. With these definitions, the energy of the Q-ball reads,
\begin{equation}
\widehat{E}=
\frac{1}{2\sin^2\varphi\,\nu}+(1-\epsilon)(1-\cos\varphi)\,\nu
+ B \left(4C^2\epsilon^2+4C\epsilon^3+\epsilon^4 \right)\,\nu
+ \frac{A}{\nu^{2/3}}\,,
\label{eq:dimless-energy-full-hpot}
\end{equation}
where we have set $\eta = 0$, as discussed above. Minimising this expression with respect to $\epsilon$, $\varphi$ and $\nu$ yields, respectively,
\begin{equation}
\label{eq:MinimisationConditions}
\begin{split}
4 B \left( 2C^2 \epsilon + 3C\epsilon^2 +  \epsilon^3 \right)& = (1 - \cos\varphi) \\
(1 - \epsilon) \nu^2 & = \frac{\cos\varphi}{(1 - \cos^{2}\varphi)^2} \\
(1-\epsilon)(1-\cos\varphi) + B \left(4C^2\epsilon^2+4C\epsilon^3+\epsilon^4 \right)
& = \frac{1}{2 (1 - \cos^{2}\varphi) \nu^2} + \frac{2}{3} \frac{A}{\nu^{5/3}}.
\end{split}
\end{equation}
It is not possible to simultaneously solve these equations analytically due to the combination of the terms proportional to $\nu^{-2}$ and $\nu^{-5/3}$, as well as the Higgs self-interactions.\footnote{One might naively expect that, in the large volume limit, we might ignore the term proportional to $\nu^2$, however, this becomes equivalent to the limits, either $\epsilon \to 1$, $\phi \to 0$, or $\nu \to 0$, which are all inconsistent with our assumptions ($\epsilon<1$) or the requirements for a stable Q-ball solution ($\phi \not= 0$ and $\nu > 0$).} We must therefore proceed numerically obtaining, at best, analytical approximations in certain limits.

Naively, we might expect that, since they represent contributions to the Q-ball energy without a corresponding contribution to the charge, the parameters $A$, $B$ and $C$ (see Eq.~\eqref{eq:dimless-pars}) should be as small as possible for a stable Q-ball to form. 
However, note that if we set $B\to0$ in the above, then $\varphi \to 0$. This represents the vacuum solution, i.e., no stable Q-ball forms. Thus, counterintuitively, the Higgs potential is a necessary component in the stabilisation of these Q-balls; it is not enough for the Higgs to merely couple linearly to the pNGBs.

In the limit of $A\to 0$ and $C \to 0$, with $B$ non-zero, these equations can be readily solved for a stable Q-ball solution. We find that, to leading order,
\begin{equation}
\varphi^2 \approx 8B\epsilon^3, \quad  \nu \approx \frac{1}{8B\epsilon^3} \quad \mathrm{and} \quad
\epsilon \approx \frac{1}{(8B)^{1/2}}.
\end{equation}
The resulting physical properties of the Q-balls are given by
\begin{equation}
m_{Q} \approx m_{\pi}Q \left( 1 - \frac{1}{8}\frac{1}{(2B)^{1/2}} \right) \quad \mathrm{and} \quad V \approx \frac{Q}{m_{\pi}f^2} (8B)^{1/2}.
\end{equation}
We see that the resulting Q-balls are stable, provided that $B> 1/128$.
In fact, $B\gtrsim\mathcal{O}(1)$ since the expansion is in terms of $\epsilon\ll 1$, and so this condition is always satisfied whenever the expansion in $\epsilon$ is valid.

For the resulting Q-balls to be well-approximated by this idealised solution, we require that $A \ll 1$ and that the Higgs potential be well-defined by its quartic term in the centre of the Q-ball, i.e., that $4C \ll \epsilon$. Considering the definition of $C$ and $\epsilon$ given above, this translates to $v_h \ll h_0$, and so the VEV of the Higgs inside the Q-ball must be much greater than its VEV outside it. This occurs if the separation between the scale of the Higgs and the scale of the HS pNGBs is large. To corroborate this, we examine this inequality in terms of the solution above,
\begin{equation}
8\sqrt{2} B^{1/2} C \ll 1,
\end{equation}
which, in terms of the fundamental parameters of the theory, is
\begin{equation}
\frac{v_s^2}{m_\pi f} \ll 1,
\end{equation}
where we have assumed that no hierarchy exists between the parameters $\lambda_s$ and $\lambda_p$, and taken $\lambda \sim 0.1$. Thus, we see that, for this limit to be realistic, we require a hierarchy between the scale of the pNGBs and the VEV of the \textit{heavier} HS Higgs. Moreover, since we ignored the HS Higgs from our analysis, for this to be self-consistent, we would require a large splitting between $m_\pi$ and $f$. We thus see that this limit is highly idealised (and is contrary to our realistic parameter values discussed above).

\begin{figure}
	\begin{centering}
		\includegraphics[width=0.49\columnwidth]{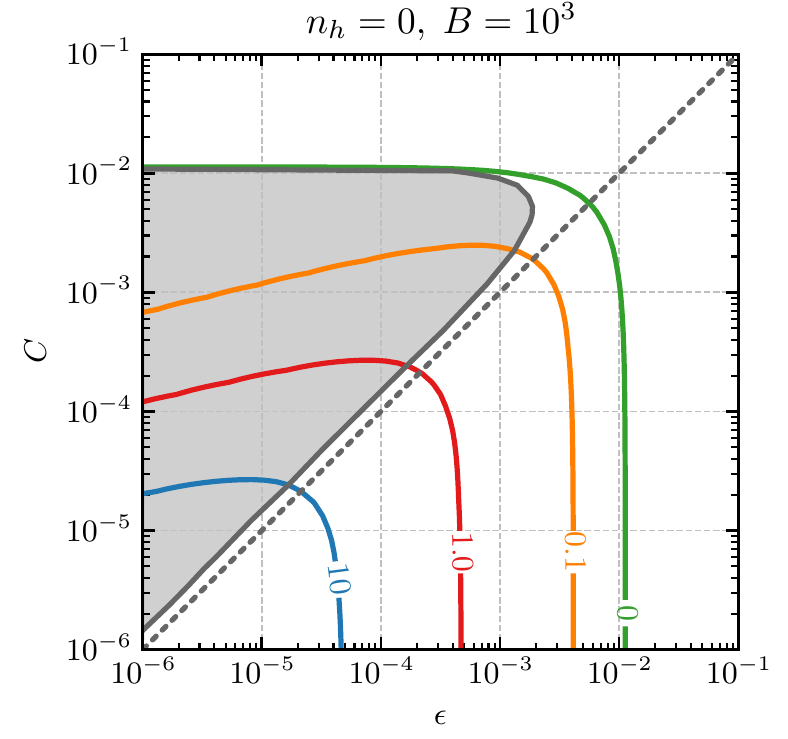}
		\includegraphics[width=0.49\columnwidth]{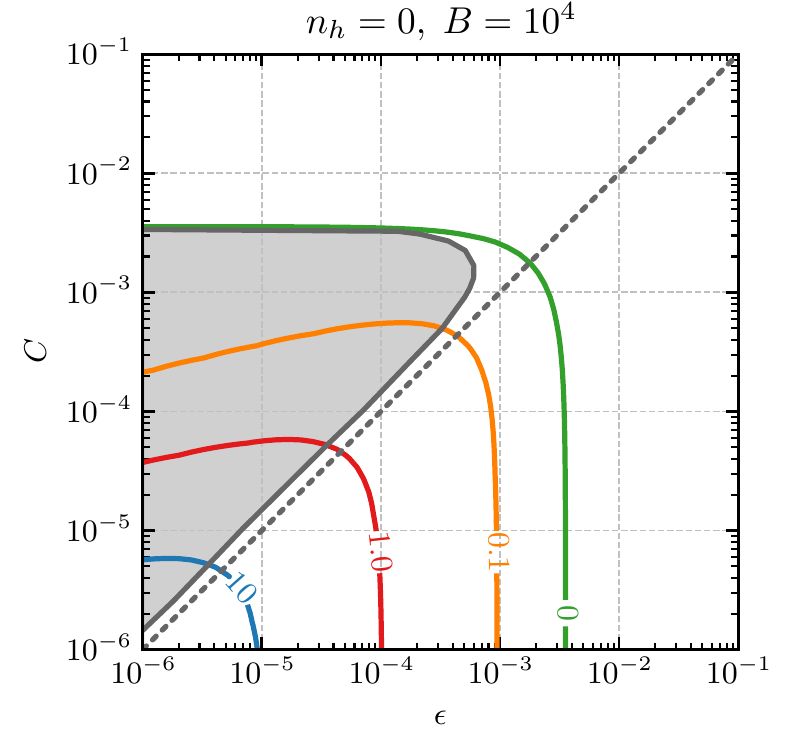}
		\caption{The $C$ vs. $\epsilon$ plane. Each line corresponds to a family of Q-ball solutions for different values of the parameter $A$ each given on their respective line. The dotted line denotes $C = \epsilon$. The excluded zone is the region where the binding energy goes negative -- this corresponds with $C\gg \epsilon$, as discussed in the main text.}
		\label{fig:c.vs.epsilon}
	\end{centering}
\end{figure}

A realistic analysis, however, must consider non-zero $A$ and $C$ (see Eq.~\eqref{eq:dimless-pars}). And, regarding the latter, for $A$ positive definite, there is a maximum value of $C$ beyond which the Q-ball becomes unbound because large values of $C$ make the contribution of the Higgs potential to the Q-ball mass large.
For a given value of $A$ and $B$, we can numerically determine the relationship between $\epsilon$ and $C$ by solving the set of equations in Eq.~\eqref{eq:MinimisationConditions} simultaneously. These contours are shown in Fig.~\ref{fig:c.vs.epsilon}. The shaded region in the figure delineates where the Q-ball solutions become unstable because $C$ becomes too large.\footnote{In this plane, this can be thought of as the region where $\epsilon \ll C$ -- this is equivalent to the limit $h_0 \ll v_h$, i.e., that the Higgs VEV inside the Q-ball is negligible. As noted above, the Higgs is a necessary component to the stability of these Q-balls, and so it is expected that this limit would not produce a meaningful solution.} Note that for the case where $A=0$, the Q-ball solutions are always bound for any value of $C$ up to its asymptotic value.

The vertical asymptotes on the right of the plot represent the case that $C \to 0$.\footnote{The asymptotes in the shaded region are unphysical, and so do not mention them further.} For $A \to 0$, this corresponds to the idealised case given above. When $C$ is appreciable w.r.t. $\epsilon$ -- when the curve differs slightly from the vertical asymptote -- we find that
\begin{equation}
\epsilon \approx \frac{1}{(8B)^{1/2}} - C.
\end{equation}
Given that $1- \hat{E} \sim \epsilon$, we thus see that the binding energy of the Q-balls is reduced, which is as expected -- though the Higgs is a necessary component for the stability of these Q-balls, the Higgs self-interactions only increase the mass without increasing the charge, and so the turning on of additional terms in the potential should always relatively reduce the binding energy. For $A \gtrsim1$, we find that the asymptotes are given by
\begin{equation}
\epsilon \approx \frac{3}{64 A B^{2/3}}.
\end{equation}
In these cases, $1- \hat{E} \sim \epsilon$ once more, and so we see that the binding energy of these Q-balls reduces quickly with increasing $A$ or, equivalently, the greater the contribution the Fermi repulsion has, the less bound the Q-ball is.

\begin{figure}
\includegraphics[width=0.49\columnwidth]{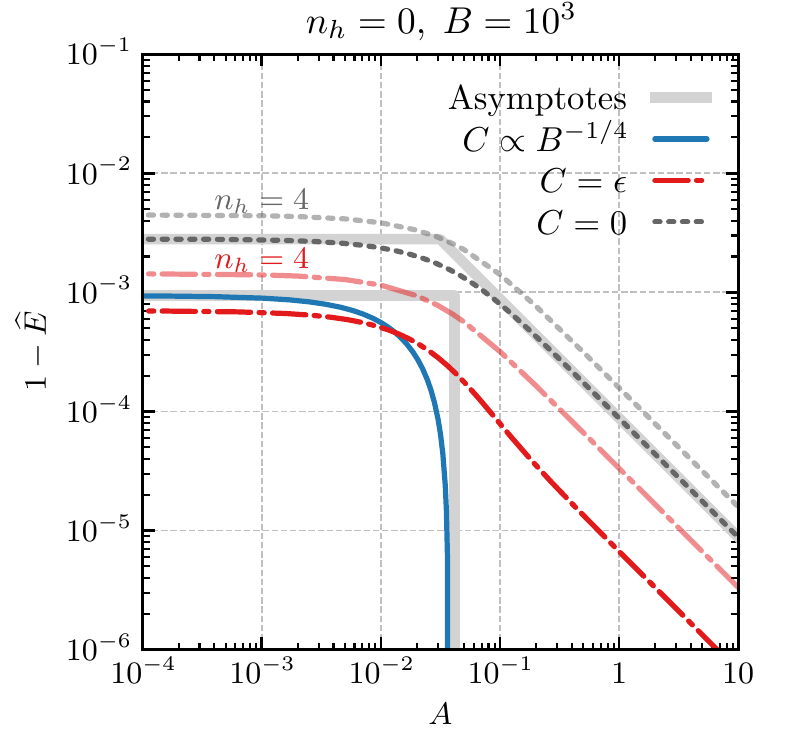}
\includegraphics[width=0.49\columnwidth]{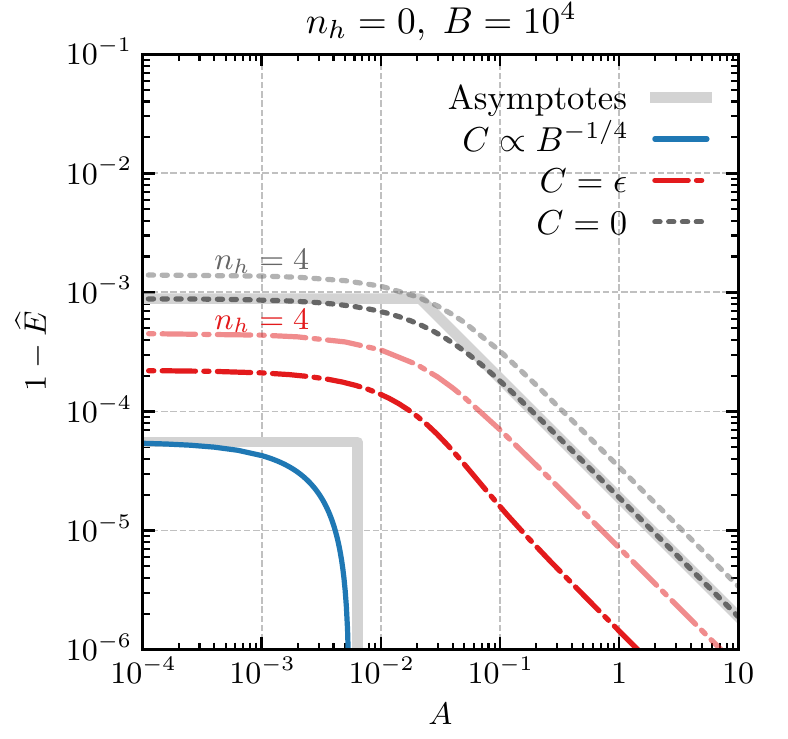}
\caption{The fractional binding energy with no heavy quarks (left panel) and three heavy quarks (right panel).  The three curves correspond to different values of the constant $B$; see text for details. The thick grey lines behind each curve (left panel) are obtained from the analytic expressions in the two limits of large and small $A$.}
\label{fig:Numerical1}
\end{figure}

\begin{figure}
\includegraphics[width=0.5\columnwidth]{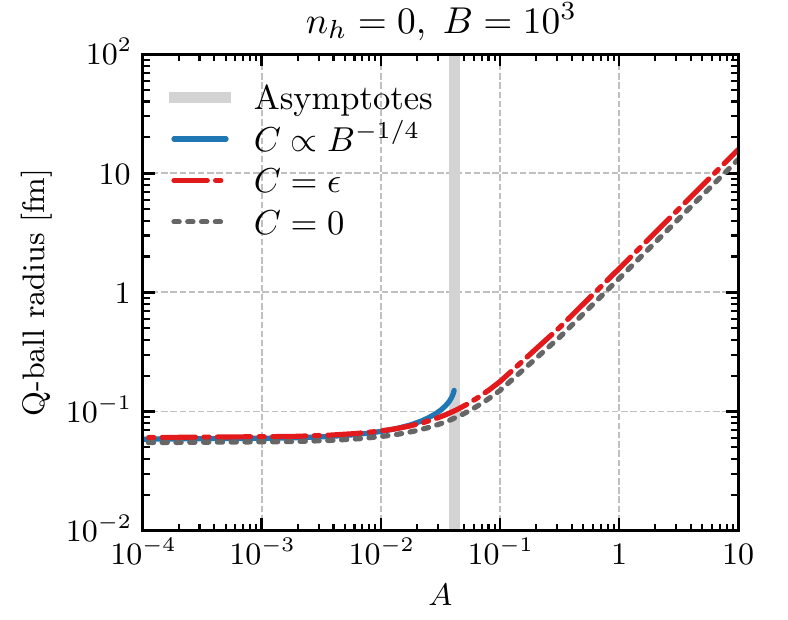}
\includegraphics[width=0.5\columnwidth]{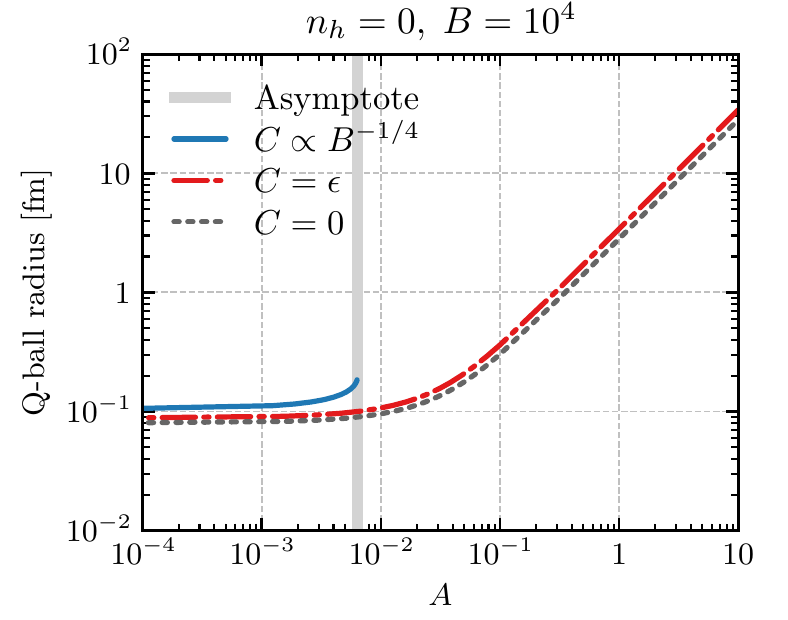}
\caption{The Q-ball radius in [fm] with $Q=10^6$. The three curves correspond to different values of the constant $B$; see text for details. The thick grey lines behind each curve (left panel) are obtained from the analytic expressions in the two limits of large and small $A$.}
\label{fig:Numerical2}
\end{figure}

We now turn our attention to the Fermi repulsion, its corresponding parameter, $A$, and the properties of realistic Q-balls. In Figs.~\ref{fig:Numerical1}~and~\ref{fig:Numerical2}, we plot the fractional binding energy, $1 - \hat{E}$, and the resulting Q-ball radius, as functions of $A$  for different values of $B$ and $C$ -- we use the fact that $B$ and $C$ are related through Eq.~\eqref{eq:BC4}. We see that there are two limiting regimes. For $A \ll 1$ -- the Fermi repulsion provides a negligible component to the energy -- there is no dependence of the physical parameters on $A$: for $C \to 0$, this regime corresponds to our idealised scenario above. In the high $A$ regime, the binding energy and radius scales as
\begin{equation}
m_{Q} \approx m_{\pi}Q \left( 1 - \frac{9}{1024}\frac{1}{AB^{2/3}} \right) \quad \mathrm{and} \quad V \approx \frac{Q}{m_{\pi}f^2} \left(\frac{32}{3}\right)^{3} A^3 B.
\end{equation}
As we can see in the latter case, together with the plots in Figs.~\ref{fig:Numerical1}~and~\ref{fig:Numerical2}, the effect of the Fermi repulsion is profound on the physical properties of the Q-ball. The binding energy reduces quickly with increased $A$ and the Q-ball radius increases quickly with increased $A$. Note, when $C \sim \epsilon$, we obtain a function of $A$ that is almost parallel to the case of $C \to 0$. When $C > \epsilon$, we see that as $A$ gets larger, it becomes too large of a component of the energy of the Q-ball, and it renders it unstable. There is thus a maximum value of $A$ that allows for stable Q-balls to form -- this maximum can only be found numerically and is model-dependent, and so we do not state any values here.

We also include in these plots the binding energy curve for non-zero heavy quarks. Here, we choose $n_h = 4$ to mimic the SM. We find that the shape of the curves are unchanged by the addition of the heavy quarks and they merely introduce a shift to the curve towards higher binding energy. This behaviour is consistent with our earlier discussion, namely that the addition of heavy quarks effectively introduces an additional, slightly larger, coupling between the HS Higgs and pions.

Finally, we comment on the phenomenology of these Q-balls if they form a component of the observed dark matter abundance. In order to give an analytic understanding, we consider the idealised scenario of $A \to 0$ and $C \to 0$ in what follows. Q-balls may not have arbitrarily large charge~\cite{Tamaki:2011zza, Sakai:2011wn, Tamaki:2011bx}. A constraint on the maximum charge of Q-balls comes from demanding that the radius is always larger than the corresponding Schwarzschild radius, $R_Q > 2 M_Q / M_{\mathrm{Pl}}^2$. This gives a constraint on the charge to be
\begin{equation}
Q < \left[\frac{3\sqrt{2}}{16\pi}\right]^{1/2} \frac{M_{\mathrm{Pl}}^3}{m_\pi^2 f} B^{1/4}.
\end{equation}
Since both the volume and energy of the Q-balls we studied scale with the charge, this in turn sets an upper bound on the mass and volume of these Q-balls before they collapse into black holes:
\begin{equation}
m_Q < \left[\frac{3\sqrt{2}}{16\pi}\right]^{1/2} \frac{M_{\mathrm{Pl}}^3}{m_\pi f} B^{1/4}
\end{equation}
and
\begin{equation}
R_Q < \frac{1}{2} \left[ \frac{54 \sqrt{2}}{\pi^3} \right]^{1/6} \frac{M_{\mathrm{Pl}}}{m_\pi f} B^{1/4} .
\end{equation}
For some typical values for the idealised case of $\mpi \sim \mathrm{TeV}$, $f \sim 10\mathrm{TeV}$, $v_s \sim 5\mathrm{TeV}$, $\theta \sim 0.01$ and $\lambda \sim 0.1$, this sets an upper bound of $\sim10$ cm! These Q-balls can therefore be phenomenologically relevant, far below their upper bound in size, in dark matter experiments seeking direct detection, as they are in principle distinguishable from point-like particle states due to form factor suppression at moderately-high momentum transfer. The question, therefore, is how can such large Q-balls be formed in the early Universe? The build-up of Q-balls from collisions with constituent quanta, coined `solitosynthesis'~\cite{Frieman:1989bx, Griest:1989bq}, is unlikely to result in such large Q-balls, or even Q-balls that are large enough to be distinguished phenomenologically. This question is the subject of ongoing study.

\section{Summary and conclusions}

In this paper, we have shown that mirror-world-like theories, with the analogue of hypercharge ungauged, can support stable Q-ball states in the thin-wall limit. The stabilisation of these Q-balls is ensured by energy and charge conservation but also requires a portal coupling with the SM Higgs. This work is an extension of previous work on the so-called thick-wall limit, valid for small charges. Here, we focussed on the thin-wall limit which describes Q-balls with large charges. We have found several regimes that allow solutions to exist, each described by the interplay between the Fermi repulsion, Higgs potential, the mixing between the two sectors, and the number of heavy ``quarks'' in the hidden sector. For a wide range of parameter values, we have shown that these solutions are stable -- however, in the case that the Higgs quadratic and cubic self-couplings are relevant, the Q-ball is rendered unstable if the Fermi repulsion is too large. In this latter case, the resulting Q-balls are sub-femtometre in size. However, if the quartic is the most relevant term in the Higgs potential, then the size of these Q-balls is primarily set by the Fermi repulsion itself and these Q-balls can be quite large in size.

In addition, if the Q-balls could contribute to the dark matter abundance inferred to exist in our Universe, the SM Higgs would provide a portal between the SM and the Q-balls. This coupling would allow for the direct detection of the Q-balls in existing dark matter experiments. Though this also requires a full phenomenological study of the evolution from formation to the detection of these objects. Such a study is warranted since the signatures in direct detection experiments would be striking. These extended objects can reach, in certain parts of parameter space, radii much greater than SM bound states, and thus the event rate can be sharply peaked at low momentum transfer due to form factor suppression. This study is the subject of future work.

\section*{Acknowledgments}

We would like to thank John March-Russell for his support, useful discussions, and comments on this work. This work was completed while OL was supported by the Science and Technologies Facilities Council (STFC) and St John's College, Oxford. This work was also supported by the Deutsche Forschungsgemeinschaft (DFG, German Research Foundation) under Germany's Excellence Strategy - EXC  2121 ``Quantum Universe'' - 390833306 and the Aspen Center for Physics, which is supported by National Science Foundation grant PHY-1607611.

\bibliographystyle{JHEP}
\bibliography{paper}

\providecommand{\href}[2]{#2}\begingroup\raggedright\begin{thebibliography}{10}

\bibitem{Lee:1991ax}
T.~D. Lee and Y.~Pang, \emph{{Nontopological solitons}},
  \href{https://doi.org/10.1016/0370-1573(92)90064-7}{\emph{Phys. Rept.}
  {\bfseries 221} (1992) 251--350}.

\bibitem{Coleman:1985ki}
S.~R. Coleman, \emph{{Q Balls}},
  \href{https://doi.org/10.1016/0550-3213(85)90286-X}{\emph{Nucl. Phys.}
  {\bfseries B262} (1985) 263}. [Erratum:
  \href{https://doi.org/10.1016/0550-3213(86)90520-1}{Nucl. Phys. B269, 744
  (1986)}].

\bibitem{Kusenko:1997zq}
A.~Kusenko, \emph{{Solitons in the supersymmetric extensions of the standard
  model}}, \href{https://doi.org/10.1016/S0370-2693(97)00584-4}{\emph{Phys.
  Lett.} {\bfseries B405} (1997) 108},
  [\href{https://arxiv.org/abs/hep-ph/9704273}{{\ttfamily hep-ph/9704273}}].

\bibitem{Safian:1987pr}
A.~M. Safian, S.~R. Coleman and M.~Axenides, \emph{{Some non-Abelian Q-balls}},
  \href{https://doi.org/10.1016/0550-3213(88)90315-X}{\emph{Nucl. Phys.}
  {\bfseries B297} (1988) 498--514}.

\bibitem{Lee:1988ag}
K.-M. Lee, J.~A. Stein-Schabes, R.~Watkins and L.~M. Widrow, \emph{{Gauged Q
  Balls}}, \href{https://doi.org/10.1103/PhysRevD.39.1665}{\emph{Phys. Rev.}
  {\bfseries D39} (1989) 1665}.

\bibitem{Heeck:2021zvk}
J.~Heeck, A.~Rajaraman, R.~Riley and C.~B. Verhaaren, \emph{{Mapping Gauged
  Q-Balls}}, \href{https://doi.org/10.1103/PhysRevD.103.116004}{\emph{Phys.
  Rev. D} {\bfseries 103} (2021) 116004},
  [\href{https://arxiv.org/abs/2103.06905}{{\ttfamily 2103.06905}}].

\bibitem{Kusenko:1997si}
A.~Kusenko and M.~E. Shaposhnikov, \emph{{Supersymmetric Q balls as dark
  matter}}, \href{https://doi.org/10.1016/S0370-2693(97)01375-0}{\emph{Phys.
  Lett.} {\bfseries B418} (1998) 46--54},
  [\href{https://arxiv.org/abs/hep-ph/9709492}{{\ttfamily hep-ph/9709492}}].

\bibitem{Kusenko:1997ad}
A.~Kusenko, \emph{{Small Q balls}},
  \href{https://doi.org/10.1016/S0370-2693(97)00582-0}{\emph{Phys. Lett.}
  {\bfseries B404} (1997) 285},
  [\href{https://arxiv.org/abs/hep-th/9704073}{{\ttfamily hep-th/9704073}}].

\bibitem{Heeck:2020bau}
J.~Heeck, A.~Rajaraman, R.~Riley and C.~B. Verhaaren, \emph{{Understanding
  Q-Balls Beyond the Thin-Wall Limit}},
  \href{https://arxiv.org/abs/2009.08462}{{\ttfamily 2009.08462}}.

\bibitem{Distler:1986ta}
J.~Distler, B.~R. Hill and D.~Spector, \emph{{$K$ Balls in the Chiral
  Lagrangian}}, \href{https://doi.org/10.1016/0370-2693(86)91080-4}{\emph{Phys.
  Lett.} {\bfseries B182} (1986) 71--74}.

\bibitem{Bishara:2017otb}
F.~Bishara, G.~Johnson, O.~Lennon and J.~March-Russell, \emph{{Higgs Assisted
  Q-balls from Pseudo-Nambu-Goldstone Bosons}},
  \href{https://doi.org/10.1007/JHEP11(2017)179}{\emph{JHEP} {\bfseries 11}
  (2017) 179}, [\href{https://arxiv.org/abs/1708.04620}{{\ttfamily
  1708.04620}}].

\bibitem{Demir:2000gj}
D.~A. Demir, \emph{{Stable Q balls from extra dimensions}},
  \href{https://doi.org/10.1016/S0370-2693(00)01262-4}{\emph{Phys. Lett. B}
  {\bfseries 495} (2000) 357--362},
  [\href{https://arxiv.org/abs/hep-ph/0006344}{{\ttfamily hep-ph/0006344}}].

\bibitem{Abel:2015tca}
S.~Abel and A.~Kehagias, \emph{{Q-branes}},
  \href{https://doi.org/10.1007/JHEP11(2015)096}{\emph{JHEP} {\bfseries 11}
  (2015) 096}, [\href{https://arxiv.org/abs/1507.04557}{{\ttfamily
  1507.04557}}].

\bibitem{Kusenko:2001vu}
A.~Kusenko and P.~J. Steinhardt, \emph{{Q ball candidates for selfinteracting
  dark matter}},
  \href{https://doi.org/10.1103/PhysRevLett.87.141301}{\emph{Phys. Rev. Lett.}
  {\bfseries 87} (2001) 141301},
  [\href{https://arxiv.org/abs/astro-ph/0106008}{{\ttfamily
  astro-ph/0106008}}].

\bibitem{Graham:2015apa}
P.~W. Graham, S.~Rajendran and J.~Varela, \emph{{Dark Matter Triggers of
  Supernovae}}, \href{https://doi.org/10.1103/PhysRevD.92.063007}{\emph{Phys.
  Rev.} {\bfseries D92} (2015) 063007},
  [\href{https://arxiv.org/abs/1505.04444}{{\ttfamily 1505.04444}}].

\bibitem{Ponton:2019hux}
E.~Pont\'on, Y.~Bai and B.~Jain, \emph{{Electroweak Symmetric Dark Matter
  Balls}}, \href{https://doi.org/10.1007/s13130-019-11194-5}{\emph{JHEP}
  {\bfseries 09} (2019) 011},
  [\href{https://arxiv.org/abs/1906.10739}{{\ttfamily 1906.10739}}].

\bibitem{Krylov:2013qe}
E.~Krylov, A.~Levin and V.~Rubakov, \emph{{Cosmological phase transition,
  baryon asymmetry and dark matter Q-balls}},
  \href{https://doi.org/10.1103/PhysRevD.87.083528}{\emph{Phys. Rev. D}
  {\bfseries 87} (2013) 083528},
  [\href{https://arxiv.org/abs/1301.0354}{{\ttfamily 1301.0354}}].

\bibitem{Gelmini:2002ez}
G.~Gelmini, A.~Kusenko and S.~Nussinov, \emph{{Experimental identification of
  nonpointlike dark matter candidates}},
  \href{https://doi.org/10.1103/PhysRevLett.89.101302}{\emph{Phys. Rev. Lett.}
  {\bfseries 89} (2002) 101302},
  [\href{https://arxiv.org/abs/hep-ph/0203179}{{\ttfamily hep-ph/0203179}}].

\bibitem{Kusenko:1997vp}
A.~Kusenko, V.~Kuzmin, M.~E. Shaposhnikov and P.~Tinyakov, \emph{{Experimental
  signatures of supersymmetric dark matter Q balls}},
  \href{https://doi.org/10.1103/PhysRevLett.80.3185}{\emph{Phys. Rev. Lett.}
  {\bfseries 80} (1998) 3185--3188},
  [\href{https://arxiv.org/abs/hep-ph/9712212}{{\ttfamily hep-ph/9712212}}].

\bibitem{Croon:2019rqu}
D.~Croon, A.~Kusenko, A.~Mazumdar and G.~White, \emph{{Solitosynthesis and
  Gravitational Waves}},
  \href{https://doi.org/10.1103/PhysRevD.101.085010}{\emph{Phys. Rev. D}
  {\bfseries 101} (2020) 085010},
  [\href{https://arxiv.org/abs/1910.09562}{{\ttfamily 1910.09562}}].

\bibitem{Kobzarev:1966qya}
I.~{\relax Yu}. Kobzarev, L.~B. Okun and I.~{\relax Ya}. Pomeranchuk, \emph{{On
  the possibility of experimental observation of mirror particles}},
  {\emph{Sov. J. Nucl. Phys.} {\bfseries 3} (1966) 837--841}. [Yad.
  Fiz.3,1154(1966)].

\bibitem{Foot:1991bp}
R.~Foot, H.~Lew and R.~R. Volkas, \emph{{A Model with fundamental improper
  space-time symmetries}},
  \href{https://doi.org/10.1016/0370-2693(91)91013-L}{\emph{Phys. Lett.}
  {\bfseries B272} (1991) 67--70}.

\bibitem{Foot:2014mia}
R.~Foot, \emph{{Mirror dark matter: Cosmology, galaxy structure and direct
  detection}}, \href{https://doi.org/10.1142/S0217751X14300130}{\emph{Int. J.
  Mod. Phys.} {\bfseries A29} (2014) 1430013},
  [\href{https://arxiv.org/abs/1401.3965}{{\ttfamily 1401.3965}}].

\bibitem{Coleman:1969sm}
S.~R. Coleman, J.~Wess and B.~Zumino, \emph{{Structure of phenomenological
  Lagrangians. 1.}},
  \href{https://doi.org/10.1103/PhysRev.177.2239}{\emph{Phys. Rev.} {\bfseries
  177} (1969) 2239--2247}.

\bibitem{Callan:1969sn}
C.~G. Callan, Jr., S.~R. Coleman, J.~Wess and B.~Zumino, \emph{{Structure of
  phenomenological Lagrangians. 2.}},
  \href{https://doi.org/10.1103/PhysRev.177.2247}{\emph{Phys. Rev.} {\bfseries
  177} (1969) 2247--2250}.

\bibitem{Voloshin:1980zf}
M.~B. Voloshin and V.~I. Zakharov, \emph{{Measuring QCD Anomalies in Hadronic
  Transitions Between Onium States}},
  \href{https://doi.org/10.1103/PhysRevLett.45.688}{\emph{Phys. Rev. Lett.}
  {\bfseries 45} (1980) 688}.

\bibitem{Voloshin:1985tc}
M.~B. Voloshin, \emph{{Once Again About the Role of Gluonic Mechanism in
  Interaction of Light Higgs Boson with Hadrons}}, {\emph{Sov. J. Nucl. Phys.}
  {\bfseries 44} (1986) 478}. [Yad. Fiz.44,738(1986)].

\bibitem{Chivukula:1989ds}
R.~S. Chivukula, A.~G. Cohen, H.~Georgi, B.~Grinstein and A.~V. Manohar,
  \emph{{Higgs decay into goldstone bosons}},
  \href{https://doi.org/10.1016/0003-4916(89)90119-X}{\emph{Annals Phys.}
  {\bfseries 192} (1989) 93--103}.

\bibitem{Chacko:2005pe}
Z.~Chacko, H.-S. Goh and R.~Harnik, \emph{{The Twin Higgs: Natural electroweak
  breaking from mirror symmetry}},
  \href{https://doi.org/10.1103/PhysRevLett.96.231802}{\emph{Phys. Rev. Lett.}
  {\bfseries 96} (2006) 231802},
  [\href{https://arxiv.org/abs/hep-ph/0506256}{{\ttfamily hep-ph/0506256}}].

\bibitem{Chacko:2005vw}
Z.~Chacko, Y.~Nomura, M.~Papucci and G.~Perez, \emph{{Natural little hierarchy
  from a partially goldstone twin Higgs}},
  \href{https://doi.org/10.1088/1126-6708/2006/01/126}{\emph{JHEP} {\bfseries
  01} (2006) 126}, [\href{https://arxiv.org/abs/hep-ph/0510273}{{\ttfamily
  hep-ph/0510273}}].

\bibitem{Chacko:2005un}
Z.~Chacko, H.-S. Goh and R.~Harnik, \emph{{A Twin Higgs model from left-right
  symmetry}}, \href{https://doi.org/10.1088/1126-6708/2006/01/108}{\emph{JHEP}
  {\bfseries 01} (2006) 108},
  [\href{https://arxiv.org/abs/hep-ph/0512088}{{\ttfamily hep-ph/0512088}}].

\bibitem{Craig:2015pha}
N.~Craig, A.~Katz, M.~Strassler and R.~Sundrum, \emph{{Naturalness in the Dark
  at the LHC}}, \href{https://doi.org/10.1007/JHEP07(2015)105}{\emph{JHEP}
  {\bfseries 07} (2015) 105},
  [\href{https://arxiv.org/abs/1501.05310}{{\ttfamily 1501.05310}}].

\bibitem{Garcia:2015loa}
I.~Garcia~Garcia, R.~Lasenby and J.~March-Russell, \emph{{Twin Higgs WIMP Dark
  Matter}}, \href{https://doi.org/10.1103/PhysRevD.92.055034}{\emph{Phys. Rev.}
  {\bfseries D92} (2015) 055034},
  [\href{https://arxiv.org/abs/1505.07109}{{\ttfamily 1505.07109}}].

\bibitem{Garcia:2015toa}
I.~Garcia~Garcia, R.~Lasenby and J.~March-Russell, \emph{{Twin Higgs Asymmetric
  Dark Matter}},
  \href{https://doi.org/10.1103/PhysRevLett.115.121801}{\emph{Phys. Rev. Lett.}
  {\bfseries 115} (2015) 121801},
  [\href{https://arxiv.org/abs/1505.07410}{{\ttfamily 1505.07410}}].

\bibitem{Craig:2015xla}
N.~Craig and A.~Katz, \emph{{The Fraternal WIMP Miracle}},
  \href{https://doi.org/10.1088/1475-7516/2015/10/054}{\emph{JCAP} {\bfseries
  1510} (2015) 054}, [\href{https://arxiv.org/abs/1505.07113}{{\ttfamily
  1505.07113}}].

\bibitem{Farina:2016ndq}
M.~Farina, A.~Monteux and C.~S. Shin, \emph{{Twin mechanism for baryon and dark
  matter asymmetries}},
  \href{https://doi.org/10.1103/PhysRevD.94.035017}{\emph{Phys. Rev.}
  {\bfseries D94} (2016) 035017},
  [\href{https://arxiv.org/abs/1604.08211}{{\ttfamily 1604.08211}}].

\bibitem{Barger:2008jx}
V.~Barger, P.~Langacker, M.~McCaskey, M.~Ramsey-Musolf and G.~Shaughnessy,
  \emph{{Complex Singlet Extension of the Standard Model}},
  \href{https://doi.org/10.1103/PhysRevD.79.015018}{\emph{Phys. Rev.}
  {\bfseries D79} (2009) 015018},
  [\href{https://arxiv.org/abs/0811.0393}{{\ttfamily 0811.0393}}].

\bibitem{Coleman:1977py}
S.~R. Coleman, \emph{{The Fate of the False Vacuum. 1. Semiclassical Theory}},
  \href{https://doi.org/10.1103/PhysRevD.15.2929}{\emph{Phys. Rev.} {\bfseries
  D15} (1977) 2929--2936}. [Erratum:
  \href{https://doi.org/10.1103/PhysRevD.16.1248}{Phys. Rev. D16, 1248
  (1977)}].

\bibitem{Callan:1977pt}
C.~G. Callan, Jr. and S.~R. Coleman, \emph{{The Fate of the False Vacuum. 2.
  First Quantum Corrections}},
  \href{https://doi.org/10.1103/PhysRevD.16.1762}{\emph{Phys. Rev.} {\bfseries
  D16} (1977) 1762--1768}.

\bibitem{Coleman:1977th}
S.~R. Coleman, V.~Glaser and A.~Martin, \emph{{Action Minima Among Solutions to
  a Class of Euclidean Scalar Field Equations}},
  \href{https://doi.org/10.1007/BF01609421}{\emph{Commun. Math. Phys.}
  {\bfseries 58} (1978) 211}.

\bibitem{Spector:1987ag}
D.~Spector, \emph{{First Order Phase Transitions in a Sector of Fixed Charge}},
  \href{https://doi.org/10.1016/0370-2693(87)90777-5}{\emph{Phys. Lett.}
  {\bfseries B194} (1987) 103}.

\bibitem{Tamaki:2011zza}
T.~Tamaki and N.~Sakai, \emph{{How does gravity save or kill Q-balls?}},
  \href{https://doi.org/10.1103/PhysRevD.83.044027}{\emph{Phys. Rev.}
  {\bfseries D83} (2011) 044027},
  [\href{https://arxiv.org/abs/1105.2932}{{\ttfamily 1105.2932}}].

\bibitem{Sakai:2011wn}
N.~Sakai and T.~Tamaki, \emph{{What happens to Q-balls if $Q$ is so large?}},
  \href{https://doi.org/10.1103/PhysRevD.85.104008}{\emph{Phys. Rev.}
  {\bfseries D85} (2012) 104008},
  [\href{https://arxiv.org/abs/1112.5559}{{\ttfamily 1112.5559}}].

\bibitem{Tamaki:2011bx}
T.~Tamaki and N.~Sakai, \emph{{What are universal features of gravitating
  Q-balls?}}, \href{https://doi.org/10.1103/PhysRevD.84.044054}{\emph{Phys.
  Rev.} {\bfseries D84} (2011) 044054},
  [\href{https://arxiv.org/abs/1108.3902}{{\ttfamily 1108.3902}}].

\bibitem{Frieman:1989bx}
J.~A. Frieman, A.~V. Olinto, M.~Gleiser and C.~Alcock, \emph{{Cosmic Evolution
  of Nontopological Solitons. 1.}},
  \href{https://doi.org/10.1103/PhysRevD.40.3241}{\emph{Phys. Rev.} {\bfseries
  D40} (1989) 3241}.

\bibitem{Griest:1989bq}
K.~Griest and E.~W. Kolb, \emph{{Solitosynthesis: Cosmological Evolution of
  Nontopological Solitons}},
  \href{https://doi.org/10.1103/PhysRevD.40.3231}{\emph{Phys. Rev.} {\bfseries
  D40} (1989) 3231}.

\end{thebibliography}\endgroup
\end{document}